\documentclass[letterpaper,twoside,british,american,prl,american,aps,showkeys,showpacs,twocolumn]{revtex4-1}
\usepackage{lmodern}

\usepackage[T1]{fontenc}
\usepackage[utf8]{inputenc}
\usepackage{xcolor}
\usepackage{pdfcolmk}
\usepackage{babel}
\usepackage{units}
\usepackage{amsmath}
\usepackage{amssymb}
\usepackage{graphicx}
\PassOptionsToPackage{normalem}{ulem}
\usepackage{ulem}
\usepackage[unicode=true,pdfusetitle,
 bookmarks=true,bookmarksnumbered=false,bookmarksopen=false,
 breaklinks=false,pdfborder={0 0 1},backref=false,colorlinks=false]
 {hyperref}

\makeatletter

\pdfpageheight\paperheight
\pdfpagewidth\paperwidth

\providecolor{lyxadded}{rgb}{0,0,1}
\providecolor{lyxdeleted}{rgb}{1,0,0}

\@ifundefined{textcolor}{}
{%
 \definecolor{BLACK}{gray}{0}
 \definecolor{WHITE}{gray}{1}
 \definecolor{RED}{rgb}{1,0,0}
 \definecolor{GREEN}{rgb}{0,1,0}
 \definecolor{BLUE}{rgb}{0,0,1}
 \definecolor{CYAN}{cmyk}{1,0,0,0}
 \definecolor{MAGENTA}{cmyk}{0,1,0,0}
 \definecolor{YELLOW}{cmyk}{0,0,1,0}
 }

\makeatother

\begin{document}

\title{Laser Induced Gas Vortices}

\author{Uri Steinitz{*}, Yehiam Prior \& Ilya Sh. Averbukh}

\affiliation{Department of Chemical Physics, Weizmann Institute of Science, 234 Herzl Street, Rehovot, Israel 76100}

\email{uri.steinitz@weizmann.ac.il}

\date{\today}
\begin{abstract}
Properly polarized and correctly timed femtosecond laser pulses have
been demonstrated to create rotational states with a preferred sense
of rotation. We show that due to conservation of angular momentum,
collisional relaxation of these unidirectionally rotating molecules
leads to the generation of macroscopic vortex flows in the gas. The
emerging velocity field converges to a self-similar Taylor vortex,
and repeated laser excitations cause a continuous stirring of the
gas.
\end{abstract}


\maketitle
Nowadays, femtosecond lasers are routinely used to manipulate and
control molecular rotation and orientation in space  \cite{Stapelfeldt2003,Friedrich1995,*Friedrich1999}.
Particularly, diverse techniques have been developed to bring gas
molecules to a rapidly spinning state with a preferred sense of rotation.
These schemes include the optical centrifuge \cite{Karczmarek1999,Villeneuve2000,Yuan2011}
and the molecular propeller \cite{Fleischer2009,Kitano2009} methods,
excitation by a chiral pulse train \cite{Zhdanovich2011} and other
techniques that are currently discussed \cite{Cryan2011,Lapert2011}.
In rarefied gas, coherent effects such as alignment revivals have
been widely investigated. Most recently, the optical centrifuge method
was applied to excite molecules to rotational levels with angular
momentum of hundreds of $\hbar$ in dense gas samples \cite{Yuan2011},
where collisions play an important role. In this paper we analyze
the behavior of a dense gas of unidirectionally spinning molecules
after many collisions have occurred. The statistical mechanics and
equilibration process of such a system are not trivial, as known for
other ensembles in which angular momentum (AM) is a conserved quantity
in addition to energy \cite{Roman2003,Uranagase2007}. We show that
due to the AM conservation, the collisions in such a gas transform
the laser induced molecular rotation into macroscopic vortex flows.
The lifetime of the generated vortex motion exceeds the typical collision
time by orders of magnitude, and the emerging velocity field eventually
converges to a self-similar Taylor vortex \cite{Taylor1918}. The
viscous decay of this whirl can be overcome by repeated laser excitations
that produce a continuous stirring effect. 

Consider a gas sample excited by laser pulses causing the molecules
to rotate, on average, unidirectionally. Our analysis of the subsequent
gas dynamics was performed in two steps. First we used a Monte Carlo
approach to simulate directly the kinetics of molecular collisions
just after the excitation. Then, after the motion had taken a collective
character, we used continuum hydrodynamic equations to investigate
the gas dynamics and vortex flow evolution.

\begin{figure}
\includegraphics[width=3.5in]{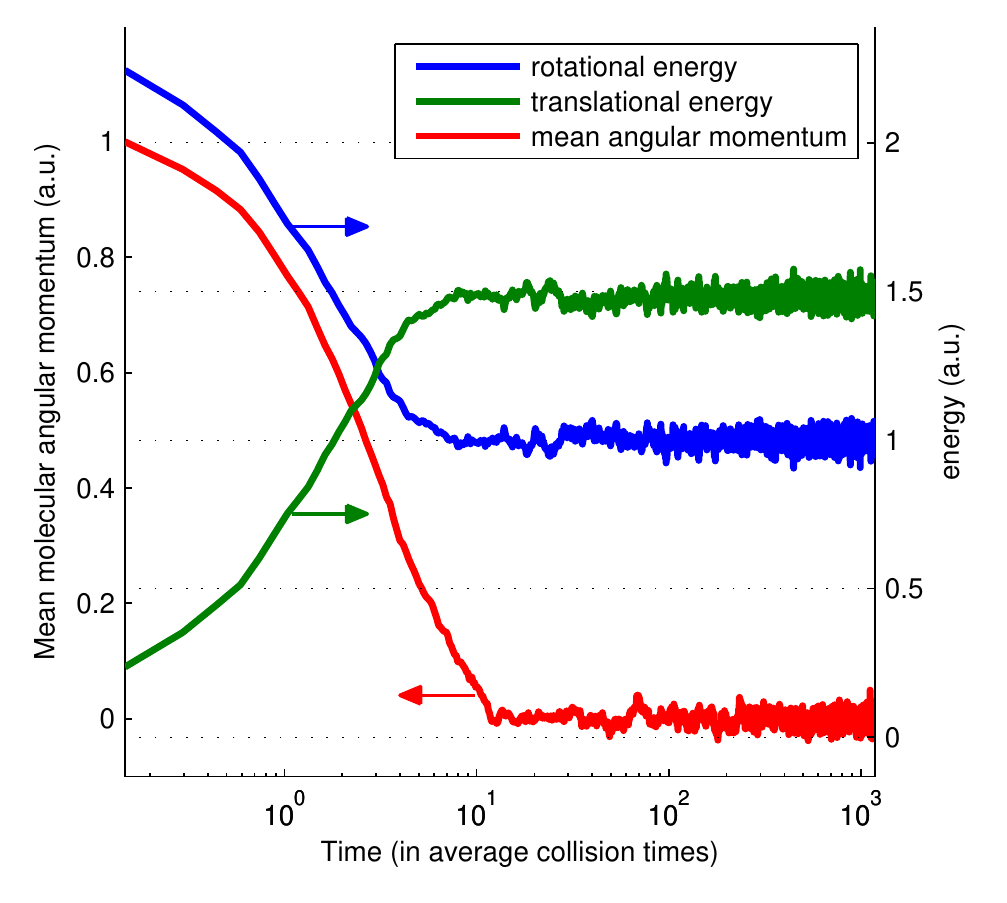}

\caption{\selectlanguage{british}%
\label{fig:energy evolution}\foreignlanguage{american}{(Color online.)
Mean translational energy, rotational energy and molecular axial rotational
angular momentum evolution. Initially all the molecules were unidirectionally
rotating. The DSMC results show that the final translational:rotational
energy ratio reaches 3:2 (green:blue curves, right axis) within about
5 collision times. The average axial component of the molecular AM
approaches zero at about 10 collisions, showing that there is no longer
a preferred rotation direction. Time axis is logarithmic in units
of average collision time ($\sim10^{-10}\, s$).}\selectlanguage{american}
}
\end{figure}

We analyzed the initial stages of the transformation of the individual
molecular rotation into collective gas motion with the help of the
Direct Simulation Monte Carlo technique (DSMC \cite{Bird1994}). This
statistical method is a proven tool for computational molecular dynamics
of transitional regimes \cite{Prasanth2008}, although the standard
DSMC does not always conserve AM \cite{Nanbu1988,Bird1987}. For this
reason, we designed an advanced variant of the DSMC simulation in
which angular momentum is strictly conserved. The simulation was run
in axisymmetric geometry, and handled collisions using a simple {}`hard
dumbbell’ model that treats each {}`molecule' as a couple of rigidly
connected hard spheres \cite{Allen1989a}. In recent experiments,
laser excitation of unidirectional rotation induced an average axial
AM in the range of $\left\langle J_{z}\right\rangle \sim10\hbar-400\hbar$
\cite{Villeneuve2000,Kitano2009}. We first simulated a homogeneous
sample of such unidirectionally rotating nitrogen molecules put at
atmospheric conditions and confined to a smooth cylinder. In order
to reduce the statistical fluctuations of the Monte Carlo simulation
we averaged the results over multiple runs \cite{Hadjiconstantinou2003}.
Figure \ref{fig:energy evolution} shows that within a few collisions
($0.3-0.5\, ns$ after excitation at our simulation conditions) the
energy partition between the translational and rotational degrees
of freedom approaches equilibrium. Despite the simplicity of our collision
model, it provides the expected energy partition ratio of 3:2 (translation:rotation)
typical for diatomic molecules. It takes about ten collision times
for the molecules to completely lose memory of the preferred direction
of their rotation, supposedly even if the molecules remain in excited
rotational states \cite{Yuan2011}.

\begin{figure}
\includegraphics[width=3.5in]{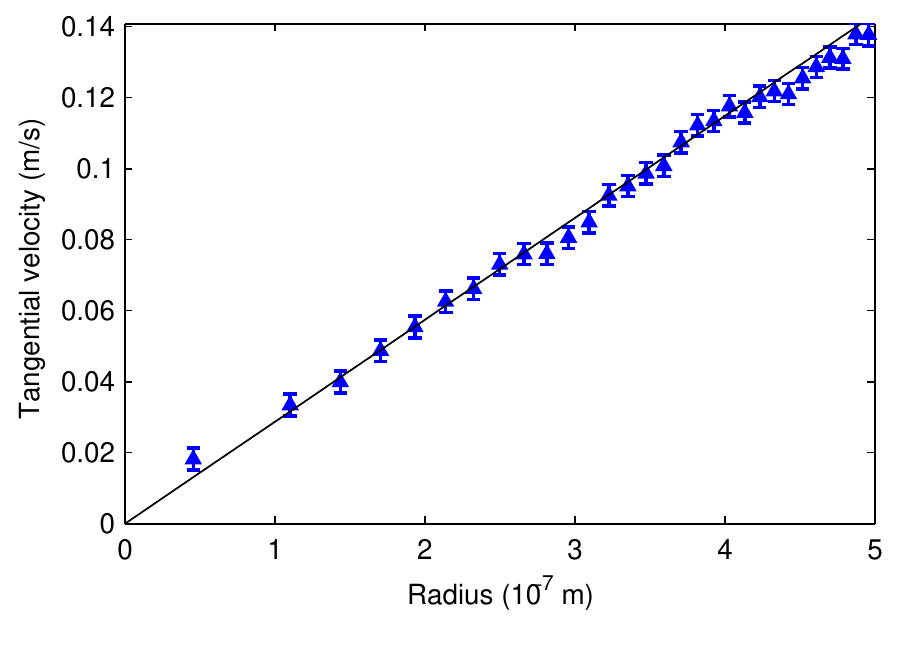}

\caption{\label{fig:tang vel vs radius}(Color online.) Tangential velocity
profile of the gas, which exhibits rigid-body-like rotation. The data
are averaged over multiple DSMC runs, covering a time range of $t=10^{-9}-2\cdot10^{-7}\, s$.
In this simulation, all the molecules initially rotate with AM of
$\left\langle J_{z}\right\rangle \sim10\hbar$. The bars indicate
the statistical error of our procedure. The solid line is a linear
fit with $R^{2}\sim0.99$\textcolor{green}{. }}
\end{figure}

At longer times, the confined gas rotates rigidly (with tangential
velocity linear with the radius) as seen in Fig. \ref{fig:tang vel vs radius}.
This represents a steady state for the motion within a frictionless
cylinder, eliminating shear stress. As expected, the results show
that the rotation speed is proportional to the total initially induced
AM, predicting a steady macroscopic gas rotation frequency of $\sim10^{4}\, Hz$
for average initial unidirectional rotation of $\left\langle J_{z}\right\rangle \sim100\hbar$
at atmospheric conditions within a $10\,\mu m$ diameter capillary. 

At the next stage we analyzed the formation of an unconfined vortex.
For this we considered excitation by an axisymmetric laser beam with
a Gaussian radial profile. Our study showed that with time, a circular
flow develops around the beam's axis. We examined the spatial and
temporal development of the circular flow structure by first running
the DSMC simulation for the non-uniform initial rotational excitation.
We sampled the results after thermalization, and used continuum gas-dynamic
analysis to unfold the subsequent gas behavior. The gas-dynamic equations
we used relate to the mass, energy and momentum conservation in a
compressible gas \cite{Colonius1991}. In order to better account
for compressibility we added bulk viscosity \cite{Gad-el-Hak1995}
terms to the equations, and the ideal gas equation of state concluded
the equation set. We used an axially symmetric finite-difference time-domain
numerical scheme to solve the problem, with initial conditions based
on the DSMC results. The full analysis of the numerical solution
will be discussed elsewhere, here we report some of our observations.

\begin{figure}
\includegraphics[width=3.5in]{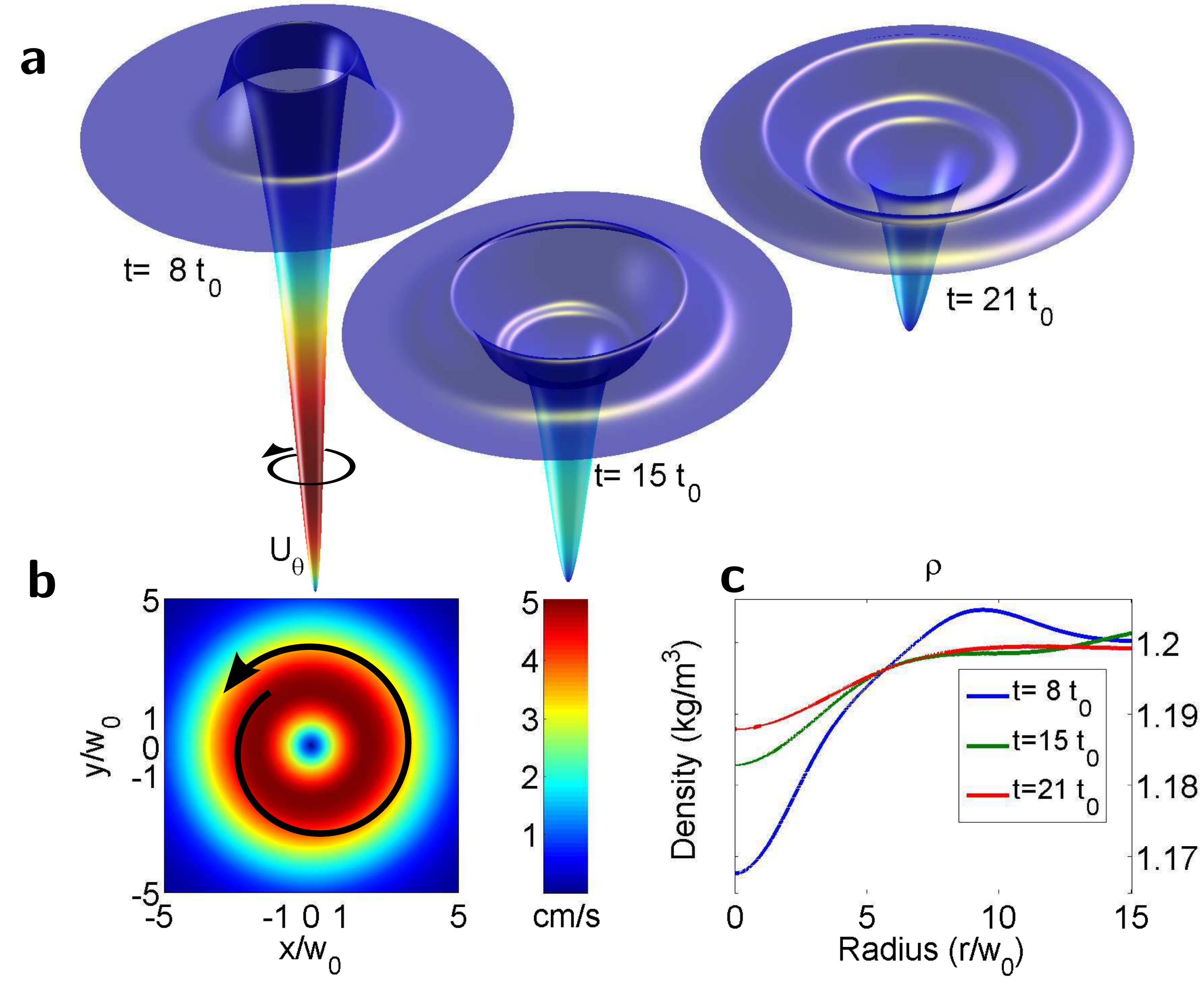}

\caption{\selectlanguage{british}%
\label{fig:vortex parametes long evolution}\foreignlanguage{american}{(Color
online.) (a) Spatial dependence of the gas density and the tangential
velocity of the unidirectional rotation at different times. The time
(from left to right) is $8\, t_{0}$, $15\, t_{0}$ and $21\, t_{0}$
after the simulation start ($t_{0}$ is the typical vortex timescale,
see Eq. \eqref{eq:time offset}). The emerging acoustic wave is clearly
visible, and the lower density core can also be noticed. The color
coding represents the tangential velocity value, also shown in (b)
in the plane perpendicular to the laser beam at $t=8\, t_{0}$. (c)
Density radial profile at the same times. The spatial coordinates
throughout are in units of $\textrm{w}_{0}$, the radius of the beam's
waist.}\selectlanguage{american}
}
\end{figure}

Since the laser injects energy to the molecules by rotating them,
the gas at the beam's focus is heated following the laser excitation.
The subsequent gas expansion creates a density crest, clearly visible
on Fig.~\ref{fig:vortex parametes long evolution}a, that propagates
outwards at about the speed of sound. 

The radial profile of the directed tangential velocity shows formation
of a rigidly rotating core at the laser beam's focal spot. The profile
exhibits a diffusion-like behavior in the radial direction. This phenomenon
is attributed to the viscosity and is consistent with the notion of
‘vorticity diffusion’ \cite{Green1995}. After the emerging acoustic
wave had left the core, naturally also the temperature spreads diffusively,
and so do the other thermodynamic and flow variables (cf. the density
plot in Fig.~\ref{fig:vortex parametes long evolution}c).

\begin{figure}
\includegraphics[width=3.5in]{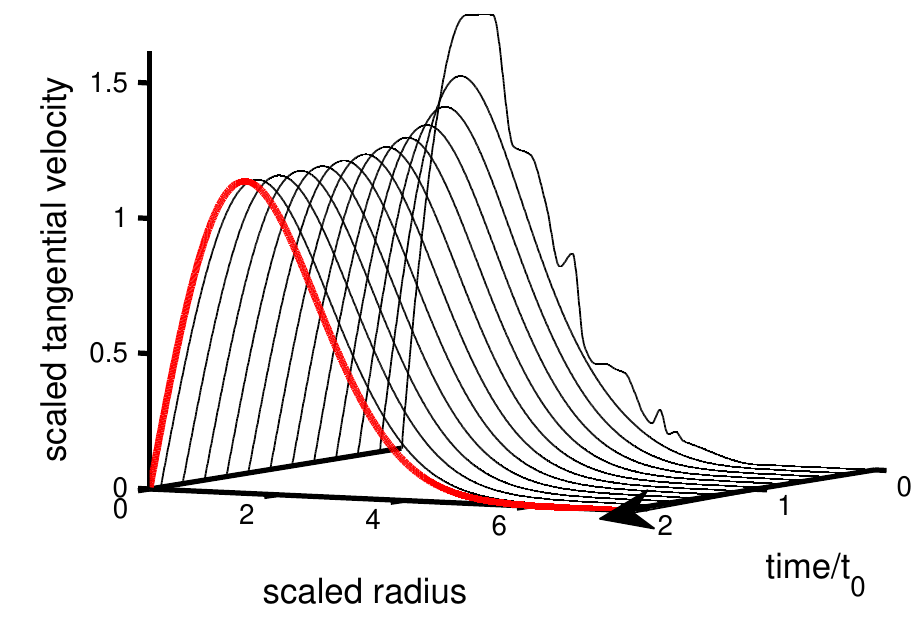}

\caption{\label{fig:Laguerre polynomial fit}(Color online.) Scaled tangential
velocity radial profile at different times. The tangential velocity
$U_{\theta}$ is scaled as $U_{\theta}t^{\nicefrac{3}{2}}\cdot\mathcal{H}^{-1}/\left[\textrm{w}_{0}t_{0}^{\nicefrac{1}{2}}\right]$,
the radius $r$ is scaled as $rt^{-\nicefrac{1}{2}}/[\textrm{w}_{0}t_{0}^{-\nicefrac{1}{2}}]$.
The plot shows a fast convergence to a self-similar Taylor vortex
profile (shaded curve, red online) given by Eqs. \eqref{eq:laguerre vorticity},\eqref{eq:time offset}.
The time $t$ was offset by a vortex {}`age' \cite{Taylor1918}
of $0.9\, t_{0}$ to produce the best fit. }
\end{figure}

At that post-acoustic stage the radial profiles of the velocity, density,
pressure and temperature at different times seem to be self-similar.
This can be illustrated by examination of the momentum conservation
equation (tangential component):

{\footnotesize 
\begin{equation}
\rho\left(\frac{\text{\ensuremath{\partial}}U_{\theta}}{\text{\ensuremath{\partial}}t}+U_{r}\frac{\partial U_{\theta}}{\text{\ensuremath{\partial}}r}+\frac{U_{r}U_{\theta}}{r}\right)=\mu\frac{\text{\ensuremath{\partial}}}{\text{\ensuremath{\partial}}r}\left(\frac{1}{r}\frac{\partial}{\partial r}\left(rU_{\theta}\right)\right)\quad,\label{eq:momentum cons:tang}
\end{equation}
}where $\rho,\, U_{\theta},\, U_{r},\mu,\, r$ and $t$ are the density,
tangential and radial velocity, viscosity, radius and time, respectively.
The velocities  (e.\,g., in Fig. \ref{fig:vortex parametes long evolution}b)
are small compared to the typical velocity ($V_{0}=\nicefrac{\mu}{\rho\textrm{w}_{0}}\sim100\,\nicefrac{m}{s}$,
$\textrm{w}_{0}$ being the inducing laser beam’s waist radius). We
may thus neglect the second order velocity terms in Eq. \eqref{eq:momentum cons:tang}.
The relative density variations also shown in Fig. \ref{fig:vortex parametes long evolution}c
are less than 10\%, and by taking a nearly constant viscosity, Eq.
\eqref{eq:momentum cons:tang} reduces to a diffusion equation, with
a {}`diffusion coefficient’ proportional to the kinematic viscosity
$\nu=\nicefrac{\mu}{\rho}$ \cite{Grasso1999}. Following that insight,
we found self-similar fits to the radial profiles of the gas parameters
(after the acoustic wave has left the core). Specifically for the
tangential velocity, Fig. \ref{fig:Laguerre polynomial fit} shows
that the scaled calculated profiles converge with time to the Taylor
vortex form \cite{Taylor1918,Neufville1957}:

\begin{equation}
U_{\theta}\left(r,t\right)=\mathcal{H}\frac{\textrm{w}_{0}}{t_{0}}\left(\frac{t}{t_{0}}\right)^{-2}2\pi\left(\frac{r}{\textrm{w}_{0}}\right)e^{-\frac{\left(\nicefrac{r}{\textrm{w}_{0}}\right)^{2}}{\left(\nicefrac{t}{t_{0}}\right)}}\quad,\label{eq:laguerre vorticity}
\end{equation}
where the length scale $\textrm{w}_{0}$ is the radius of the waist
of the laser beam, and the typical vortex lifetime is determined by
$t_{0}$. The dimensionless parameter $\mathcal{H}$ depends on the
average induced AM, and characterizes the number of turnovers the
core makes before the vortex’ viscous decay. The parameters $t_{0}$
and $\mathcal{H}$ are defined as

\begin{equation}
t_{0}=\frac{\textrm{w}_{0}^{2}}{4\nu},\quad\mathcal{H}=\frac{\left\langle J_{z}\right\rangle }{8\pi m_{mol}\cdot\nu}\quad,\label{eq:time offset}
\end{equation}
where $\left\langle J_{z}\right\rangle $ and $m_{mol}$ are the average
AM of the molecules excited by the laser and the mass of the individual
molecules, respectively. 

It follows from Eq. \eqref{eq:laguerre vorticity} that the rigid
core of the vortex rotates at a frequency that drops with time as
$t^{-2}$, while the typical core size grows as $t^{\nicefrac{1}{2}}$.
For a tightly focused laser beam ($\textrm{w}_{0}\sim1\,\mu m$) that
excites nitrogen molecules at atmospheric conditions to $\left\langle J_{z}\right\rangle \sim100\hbar$,
the vortex parameters are $\mathcal{H}\sim10^{-2},\, t_{0}\sim10^{-8}\, s$.
In this case a substantial initial rotation frequency of $\sim10^{5}\, Hz$
may be achieved.

\begin{figure}
\includegraphics{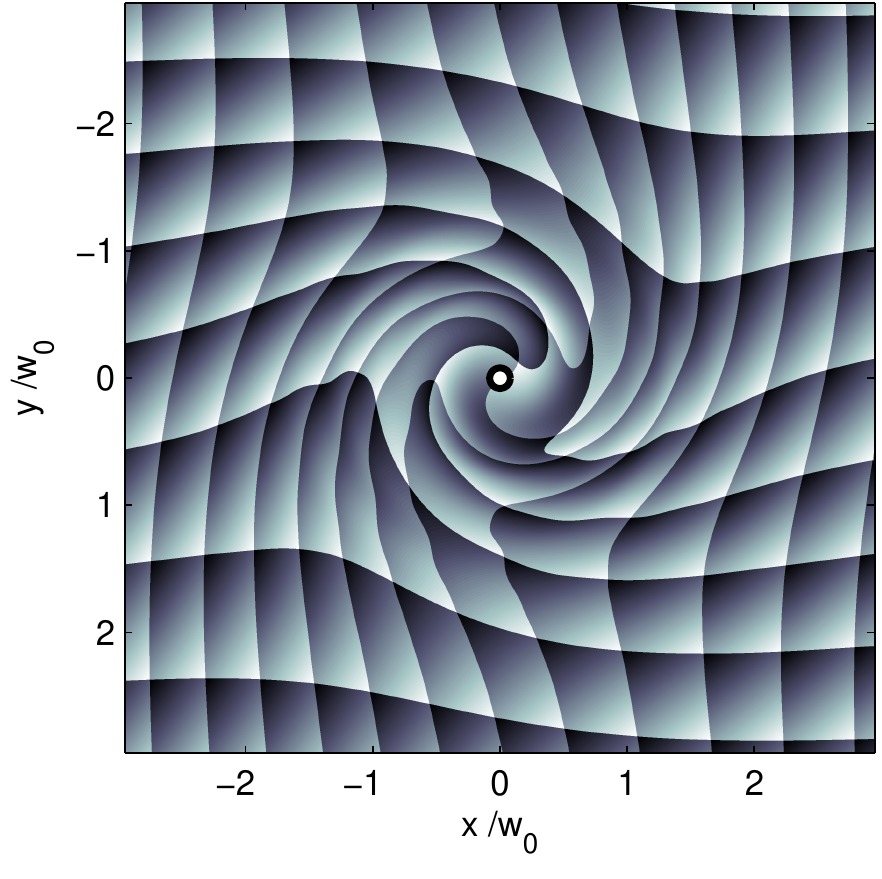}\caption{\selectlanguage{british}%
\label{fig:Total-displacement-deformation}\foreignlanguage{american}{Accumulated
displacement deformation of a virtual rectangular mesh after 6,000
pulse sets, for $\mathcal{H}\sim10^{-4}$. The angular displacement
is clearly visible around the center (circle marker). The coordinates
are in the plane perpendicular to the laser beam, in units of the
laser beam's waist radius $\textrm{w}_{0}$.}\selectlanguage{american}
}
\end{figure}

Consider now the motion of gas parcels as the vortex develops and
decays. We calculated the parcels' trajectories both numerically
and analytically with the help of Eq. \eqref{eq:laguerre vorticity},
mapping each point to its location after the vortex had died.\emph{
}The jolt imparted by the laser excitation results in a finite angular
displacement of the gas parcel. By repeating the excitation again
and again the angular displacement accumulates, rotating parcels around
the beam's axis as a {}`micro-torque-gun’, producing a motion similar
to that of a ticking watch dial. At a pulse repetition rate of $1\, kHz$,
the time interval between pulses is long enough for diffusion to wipe
the temperature and density inhomogeneities and any radial displacement,
even on millimeter-scale. Thus, in the micron-size waist of the laser
beam only the azimuthal displacement grows from jolt to jolt. Figure
\ref{fig:Total-displacement-deformation} and the Supplementary Video
\cite{PRL_SM_vortices_2012} visualize the angular displacement after
many iterations. This angular displacement pattern can be observed
by following the motion of micron-sized particles suspended in the
whirling gas.

To conclude, we showed that laser pulses that induce unidirectional
rotation of individual gas-molecules produce a long-lasting macroscopic
rotation of the gas. In a frictionless cylinder the gas reaches a
steady rigid-body-like rotation. In free space a self-similar vortex
forms and lasts significantly longer than both the pulse duration
and the typical collision time. We also demonstrated that a repeated
laser excitation can stir the gas in a pounding `torque-gun` motion
accumulating to large angular displacement. The $100\, kHz$ rotational
motion of the gas vortex may be detected with the help of the rotational
Doppler effect \cite{Garetz1981a,Allen1994,Barreiro2006} or by usage
of structured light spectroscopy \cite{Andrews2008}. According to
our estimations the predicted collective spiraling motion is also
observable by monitoring the motion of particles suspended in the
gas. Finally, we emphasize that the appearance of the gas vortex is
a direct manifestation of angular momentum conservation. In a sense,
our laser-induced effect presents a table-top analogue to the formation
of tropospheric cyclones, in which global rotational structures appear
due to the inverse cascade scale-up starting from the motion of turbulent
eddies \cite{Fedorovich2004,Tabeling2002}. 
\begin{acknowledgments}
Financial support for this research from the Israel Science Foundation
is gratefully acknowledged. This research is made possible in part
by the historic generosity of the Harold Perlman Family. The authors
thank Dennis C. Rapaport, Robert J. Gordon and Rigoberto Hernandez
for helpful discussions.
\end{acknowledgments}
\bibliographystyle{apsrev4-1}

\end{document}